\documentclass[a4paper,10pt]{article}
\usepackage[utf8]{inputenc}
\usepackage{graphicx}
\title{ Entropy spectrum of (1+1) dimensional stringy black holes}
\author{Jishnu Suresh$^{1}$ and V. C. Kuriakose$^2$ \\
\vspace{0.2in} \\
 Department of Physics, \\
Cochin University of Science and Technology, 
\\{Kochi-22, India}.\\
\vspace{0.1in} \\
E-mail: $^1$ jishnusuresh@cusat.ac.in \\
$^2$vck@cusat.ac.in\\}
\begin{document}

\maketitle
\begin{abstract}
We explore the entropy spectrum of $(1+1)$ dimensional dilatonic stringy black holes via the adiabatic
invariant integral method and the Bohr-Sommerfeld quantization rule. It is found that the corresponding spectrum depends
on black hole parameters like charge, ADM mass and more interestingly on the dilatonic field. We calculate the entropy 
of the present black hole system via the Euclidean treatment of quantum gravity and study the thermodynamics
of the black hole and find that the system does not undergo any phase transition.
\end{abstract}

\section{Introduction}

The fundamental difficulties in merging quantum theory with gravitation theory are well known. We know that the beauty of black hole thermodynamics
lies in the powerful way it speaks of the unity of physics. To find answers to the difficulties, plenty of studies are going on 
by applying the known laws of quantum mechanics to General Relativity. As a result of these studies, it is found that it is very difficult
to study the systems in four dimensions. So enormous studies have been done on lower dimensional gravity theories. String theory is one approach 
to quantum gravity. The developments in string theory have provided a good framework to consider the quantum properties of the black holes. The low 
energy string theory has several black hole solutions \cite{witten,teo,nappi}. The study of lower dimensional black hole systems 
will help us to address many problems that
arise in higher dimensional quantum gravity models. Hence several studies have been done in black hole thermodynamics of two dimensional gravity models. 
In addition, from these understandings as well as using black hole/string correspondence principle one can understand the conceptual issues regarding the 
microscopic origin of black hole entropy.

In recent years the dilatonic black holes have gained much attention, because it is widely believed that, the investigations on this black 
hole solution will lead to an exact explanation of microscopic origin of black hole entropy (Bekenstein-Hawking entropy) \cite{myers,sadeghi,hyun}.
Also there exists some conceptual problems 
regarding the end point of black hole evaporation through thermal radiation \cite{kim,vagenas1,easson}. The two dimensional 
Einstein-Hilbert action is just a Gauss-Bonnet term, a topological invariant in the two dimensional space time.  So the two dimensional models are 
locally trivial. Hence it is necessary to introduce some extra fields to this model. The best candidate for this is the dilaton field, which arises in the 
compactifications from higher dimensional models or from string theory.  This theory also possesses different black hole solutions. These solutions 
play an important role in our understanding of quantum gravity. Now, as pointed out in \cite{uduality}, higher dimensional black holes in string theory
can be related 
to two dimensional solutions through U-duality. Hence the higher dimensional black holes in string theory \cite{teo,cardoso1,cardoso2,kiem,cadoni}
are related to many two dimensional solutions 
including the two dimensional charged black holes of McGuigan, Nappi and Yost \cite{nappi}. 
In the present study the two dimensional dilatonic black hole is considered
which is analogous to the above mentioned two dimensional string black hole.

Bekenstein \cite{bekestein1} proposed that the horizon area of a non extremal black hole
is an adiabatic invariant, quantities which vary very slowly compared to the variation of the external perturbation of the 
system, and the horizon area of black hole is quantized in units of
Planck length. As a result of this, much attention has been given to the entropy spectrum 
quantization. Many methods have been put forward to calculate this entropy spectrum spacing. First method in this 
direction was introduced by Hod \cite{hod1,hod2}. This method relies on quasinormal modes and the corresponding frequency. Hod employed 
Bohr's correspondence principle to quantize the entropy and found that in the asymptotic limit, it is related to the real part of quasinormal
frequencies. Later Kunstatter \cite{kunstatter} derived an equally spaced entropy spectrum for Schwarzschild black hole. For this derivation, the quantity 
$I=\int{\frac{dE}{\omega_R}}$ was taken as the adiabatic invariant, where $E$ is the energy and $\omega_R$ is the real part of the 
quasinormal frequency. Maggiore in 2007 \cite{maggiore} refined the idea of Hod by proving that the physical frequency of the quasinormal modes 
is determined by its real and imaginary parts. Recently Majhi and Vagenas \cite{majhi} proposed a new method to quantize the
entropy without using quasinormal modes. They used the idea of connecting adiabatic invariant quantity 
to the Hamiltonian of the black hole, and they obtained an equally spaced entropy spectrum. Jiang and Han \cite{jiang} modified this 
idea by pointing the fact that, the adiabatic invariant quantity $\int{p_i dq_i}$ used in Majhi and Vagenas's method is not canonically invariant. 
Therefore they made the modification by using the adiabatic invariant quantity of the covariant form $I=\oint{p_i dq_i}$. We adopt this method to
quantize the entropy of the two dimensional stringy black hole.

The organization of this paper is as follows. In Sec.\ref{thermo}, we introduce the two dimensional charged dilatonic black hole solution
and discuss the thermodynamics of the corresponding black hole. In Sec.\ref{entropyformalism}, we obtain the entropy  of the black hole via 
Euclidean treatment of quantum gravity and also the entropy spectrum and the corresponding spacing is studied. Paper concludes with a short discussion
of the results in Sec.\ref{conclusion}

\section{Two dimensional-stringy black hole and its thermodynamics}
\label{thermo}
The action corresponding to Maxwell-gravity coupled to a
dilatonic field ($\Phi$) can be described by \cite{nappi}
\begin{equation}
S=\frac{1}{2\pi }\int d^{2}x\sqrt{-g}e^{-2\Phi }\left( R+4(\nabla \Phi
)^{2}-\lambda -\frac{1}{4}F_{\mu \nu }F^{\mu \nu }\right)
\label{action}
\end{equation}
in which $R$ is the Ricci scalar, $\lambda$ is the effective central charge
and $F_{\mu\nu}$ is the electromagnetic field tensor. The equations of motion corresponding 
to the metric, gauge and dilaton fields are respectively given by,
 \begin{eqnarray}
  R_{\mu \nu }-2\nabla _{\mu }\nabla _{\nu }\Phi -\frac{1}{2}F_{\mu \sigma
  }F_{\nu }^{\sigma } &=&0~,  \nonumber \\
 \nonumber \nabla _{\nu }\left( e^{-2\Phi }F^{\mu \nu }\right)  &=&0~, \\
  R-4\nabla _{\mu }\nabla ^{\mu }\Phi +4\nabla _{\mu }\Phi \nabla ^{\mu }\Phi
  -\lambda -\frac{1}{4}F_{\mu \nu }F^{\mu \nu } &=&0~. 
 \label{eqnmotions}
 \end{eqnarray}
The solution to the equation of motion leads to the two dimensional dilatonic black hole, whose 
metric is given by,
\begin{equation}
 ds^{2}=-f(r)dt ^{2}+\frac{dr^{2}}{f(r)}~, 
\label{metric1}
\end{equation}
with the metric function,
\begin{equation}
 f(r)=1-2me^{-Qr}+q^{2}e^{-2Qr}~,
\label{fofr}
\end{equation}
the dilaton field,
\begin{equation}
 \Phi =\Phi _{0}-\frac{Q}{2}r~,
\label{dilatonfield} 
\end{equation}
and the electromagnetic field tensor as,
\begin{equation}
 F_{tr}=\sqrt{2}Qqe^{-Qr}~.
\label{emstrengthtensor} 
\end{equation}
The condition of asymptotic flatness for the spacetime requires $\lambda=-Q^2$. In these equations $m$ and $Q$ are proportional to black hole mass and
black hole charge respectively. Then the horizons are located at,
\begin{equation}
r_{\pm }=\frac{1}{Q}\ln \left( m\pm \sqrt{m^{2}-q^{2}}\right)~,
\label{horizonradius}
\end{equation}
From the above expression it is evident that, in order to have an event horizon at $r_{\pm}$, 
the condition $m^2 - q^2 \geq 0$ has to be satisfied. The solution given by (\ref{fofr}) is analogous to the string theoretic black hole \cite{witten,teo,nappi}. 
So we are calling this particular black hole solution as two dimensional charged "stringy" black holes.

\begin{figure}[h]
 \includegraphics[width=1\columnwidth]{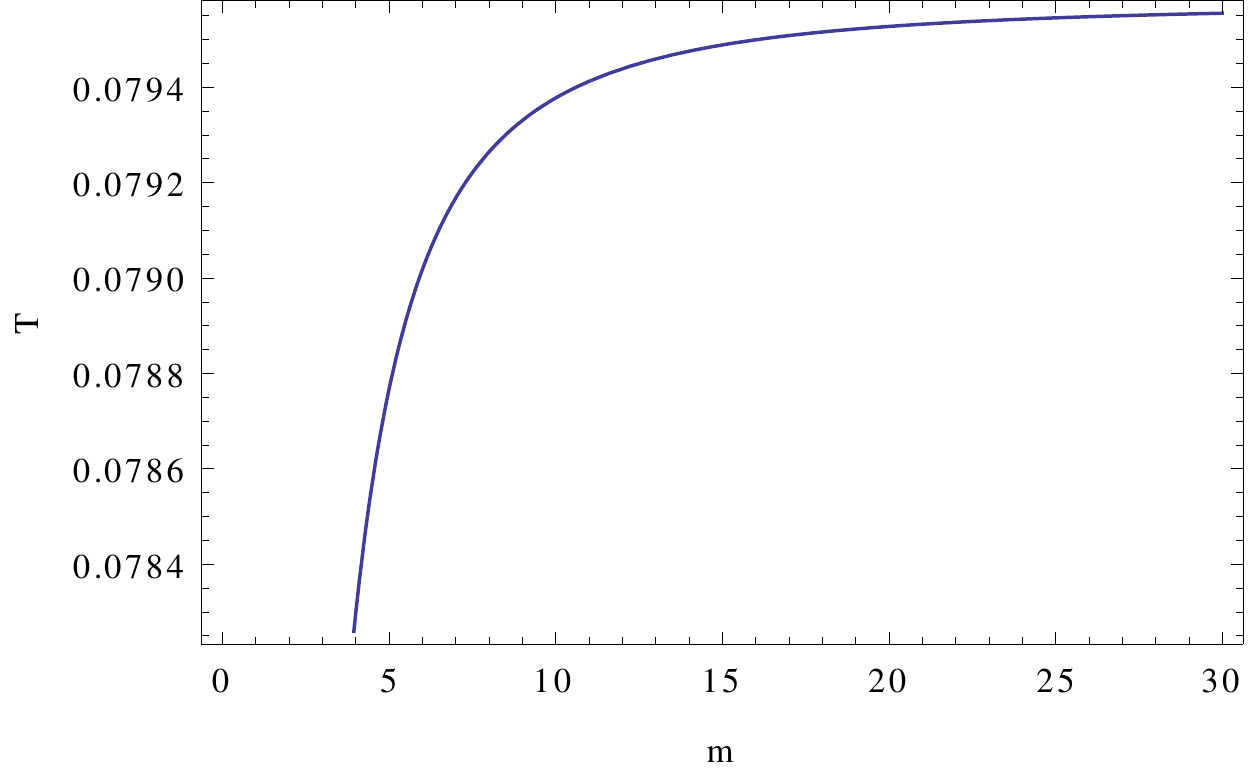}
 \caption{Variation of temperature of the black hole with respect to mass of the black hole in Planck units.}
 \label{temp_stringy}
\end{figure}
\begin{figure}[h]
 \includegraphics[width=1\columnwidth]{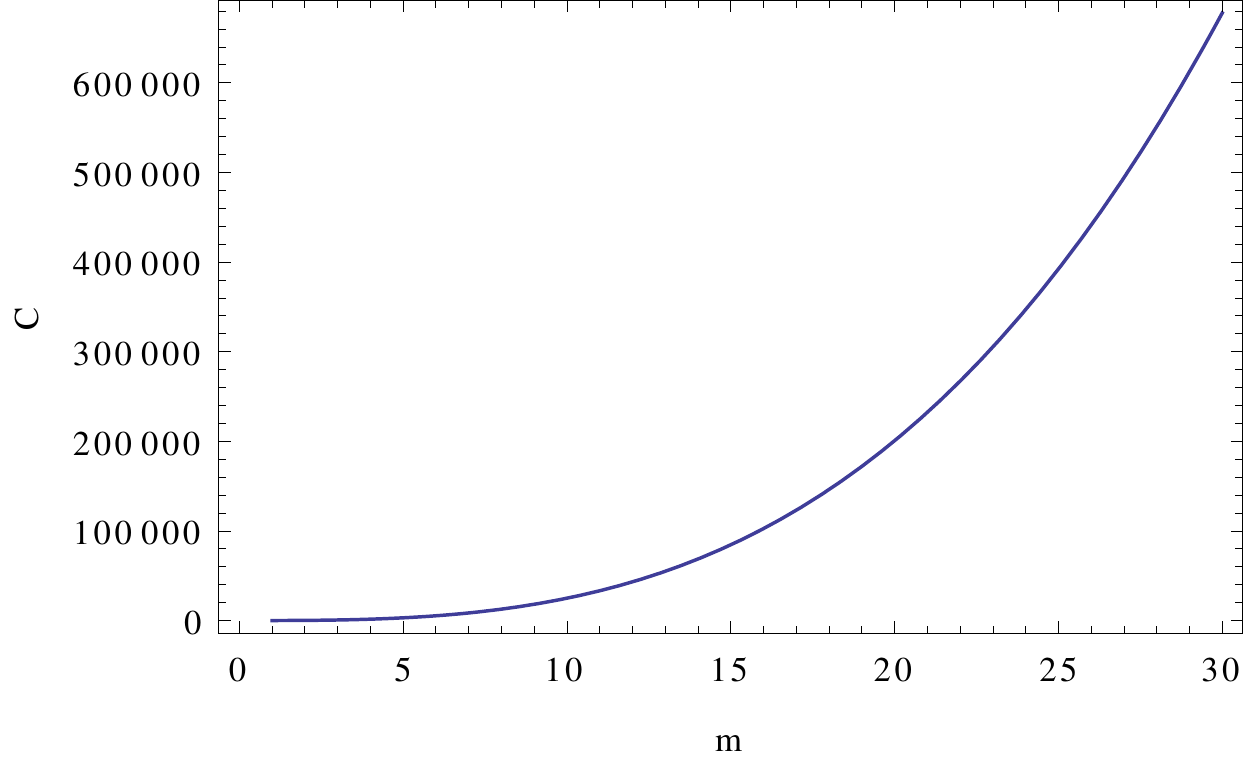}
 \caption{Variation of specific heat of the black hole with respect to mass of the black hole in Planck units.}
 \label{spec_stringy}
\end{figure}

Now we will investigate the thermodynamic aspects of charged stringy black hole. 
From (\ref{fofr}), and using the condition $f(r)=0$, at the horizon, we can deduce the mass of the black hole as,
\begin{equation}
 m= \frac{e^{Qr}+q^{2} e^{-Qr} }{2}~.
 \label{massm}
\end{equation}
The temperature of the two dimensional charged dilatonic black hole or stringy black hole
can be derived using the relation,
\begin{equation}
 T_H=\frac{\kappa}{2 \pi} 
\end{equation}
where the surface gravity $\kappa$ is defined as,
\begin{equation}
 \kappa= \frac{1}{2} \frac{\partial f(r)}{\partial r} \vline_{r=r_+} ~.
\end{equation}
This yields the Hawking temperature as
\begin{equation}
 T_H=\frac{\lambda \sqrt{m^2-q^2}}{\pi(m+\sqrt{m^2-q^2})}~,
 \label{hawking temperature_stringy}
\end{equation}
and when the charge of the string becomes zero, the temperature will reduce to
\begin{equation}
 T=\frac{\lambda}{2 \pi}~.
\end{equation}
The variation of Hawking temperature with respect to mass of the black hole is depicted in fig.(\ref{temp_stringy}). 
From the thermodynamic relation, $C=\frac{\partial m}{\partial T}$, one can arrive at the specific heat of the black 
hole as,
\begin{equation}
 C=\frac{2 \pi  \left(m \sqrt{m^2-q^2}+m^2-q^2\right)}{Q \left(m-\sqrt{m^2-q^2}\right)}~,
 \label{specific heat_stringy}
\end{equation}
and the corresponding variation with respect to mass of the black hole is plotted in fig.(\ref{spec_stringy}).
From these two figures it is evident that the two dimensional dilatonic stringy black
hole does not undergo any kind of phase transition. Ernst Ising solved the 1D Ising model in 1925 and found that there is no phase transition in
1D systems. We can see that these results match with the $(1+1)$ dimensional dilatonic stringy black hole system. This system also shows no phase 
transition. So this black hole behaves like a 1D Ising model system as far as thermodynamic behaviours are considered.

\section{Entropy and entropy spectrum}
\label{entropyformalism}
As we mentioned in the introduction, the best way to study the thermodynamics of gravitational
field is via the Euclidean treatment of quantum gravity. The partition function 
that explains the black hole thermodynamics is evaluated as the Euclidean path integral over the space of all
field configurations with the saddle point 
approximation around the black hole solution \cite{gibbons}. It is found that the entropy of the black hole under consideration
is originated from the above mentioned Euclidean path integral calculations. By following similar arguments one can perform
an analysis to obtain the entropy of a two dimensional dilatonic stringy black hole.  This analysis can begin by 
writing the partition function as,
\begin{equation}
 Z= e^{I_E}=e^{\beta F},
 \label{partition_function}
\end{equation}
where $I$ is the action evaluated for the Euclideanized gravitational field, $\beta$ is the inverse temperature, and $F$ is the 
Helmholtz free energy, $F=M-TS$. So this equation implies that the Euclidean action evaluated on the black hole system can be identified as
inverse temperature times the free energy of the system.
Here the Euclidean action is given by,
\begin{equation}
 I_E=\beta M - \frac{S}{\kappa_B}+\beta \sum_i \mu_i Q_i,
 \label{euclidean_action}
\end{equation}
where $\mu_i$'s are the chemical potentials which correspond to $Q_i$'s, which are the charges. From the
Wick-rotated form of (\ref{action}), the Euclidean minisuperspace action can be constructed. From this
construction the corresponding black hole entropy can be calculated by comparing it with the partition function and it is given by 
\begin{equation}
 S=4\pi e^{-2 \Phi_0} (m+\sqrt{m^2-q^2}).
 \label{entropy_stringy}
\end{equation}
Now it can be also possible to write down the ADM mass of the black hole as described in \cite{witten} as,
\begin{equation}
 M=2mQe^{-2\Phi_0}.
 \label{adm mass}
\end{equation}
This mass relation can also be derived by adopting the derivation of Arnowitt, Deser and Misner \cite{adm}. It is also found that this
ADM mass agrees with the thermodynamic evolution of energy. Now one can arrive at an expression, where $S$ as a function
of ADM mass ($M$) and electric charge ($Q_{el}$) as follows,
\begin{equation}
 S = \frac{2\pi}{Q}( M + \sqrt{M^2-2Q^2_{el}}),
 \label{entropyredefine}
\end{equation}
where the electric charge is given by,
\begin{equation}
 Q_{el}=\sqrt{2} q Q e^{-2 \Phi_0} = \frac{q}{\sqrt{2}m} M .
\end{equation}

Obviously one can expect a matching between this result and that of near extremal $AdS_2$ type black holes as considered in \cite{nappi}.
The $AdS_2$ type black hole solution is given by,
\begin{equation}
 g(r)=1-(m/\lambda)e^{-2 \lambda r} + (q^2 / 4 \lambda^2) e^{-4 \lambda r}.
\end{equation}
Now let us consider the near horizon limit of extremal black holes in above equation using $m=q$ condition. In this case we can find the 
black hole entropy as,
\begin{equation}
 S=\frac{\sqrt{2} \pi q}{\lambda}.
 \label{entropyredefineads}
\end{equation}

Now it can be shown from (\ref{entropyredefine}) that the entropy of the flat dilatonic black hole agrees with the entropy (\ref{entropyredefineads})
of the near extremal limit of $AdS_2$ type black hole solution \cite{nappi,hyun,perry}.
  
Now we will quantize the entropy of the two dimensional dilatonic stringy black hole via the adiabatic invariance and Bohr-Sommerfeld 
quantization rule. According to the tunneling picture, when a particle tunnels in or out, the black hole
horizon will oscillate due to the gain or loss of the black hole mass \cite{parikh}. Such oscillating horizon can be studied using the adiabatic 
invariant quantity,
\begin{equation}
 I=\oint p_i d q_i=\int_{q_{i} ^{in}} ^{q_{i} ^{out}} p_i ^{out} d q_i+\int_{q_{i} ^{out}} ^{q_{i} ^{in}} p_i ^{in} d q_i~,
 \label{adiabatic covariant action}
\end{equation}
where $p_i ^{in}$ or $p_i ^{out}$ is the conjugate momentum
corresponding to the coordinate $q_i ^{in}$ or $q_{i} ^{out}$,
respectively, and $i= 0, 1, 2..~$. For the horizon of a black hole 
$q^{in}_{1}=r^{in}_{h}(q^{out}_{1}=r^{out}_{h})$ and $q_0 ^{in}
\left( q_0 ^{out} \right)=\tau$ where $\tau$ is the Euclidean time and $r_h$ is the horizon radius.
By implementing the Hamilton's equation $\dot q_i = \frac{dH}{dp_i}$, where $H$ is the Hamilton of the system, the 
integral can be rewritten as,

\begin{eqnarray}
\int_{q_{i} ^{out}} ^{q_{i} ^{in}} p_i ^{in} d q_i &=& \int_{\tau_{out}} ^{\tau_{in}} \int_{0} ^{H} dH^\prime d\tau
 +\int_{r_{h} ^{out}} ^{r_{h} ^{in}} \int_{0} ^{H} \frac{dH^\prime}{\dot r_{h}} dr_h \nonumber \\
  &=&2\int_{r_{h} ^{out}} ^{r_{h} ^{in}} \int_{0} ^{H} \frac{dH^\prime}{\dot r_{h}} dr_h~ ,
 \label{eq:wideeq}
\end{eqnarray}

where $\dot r_h= \frac{dr_h}{d\tau}$ is the oscillating velocity of the black hole horizon. 
We know that the tunneling and oscillation take
place at the same time. Hence we can write the relation connecting the black hole horizon oscillating velocity and velocity 
of the tunneling particle as \cite{zhang},
\begin{equation}\label{oscillating velocity}
 \dot r_{h}= -\dot r ~.
\end{equation}
The metric given by (\ref{metric1}) is Euclideanized by introducing the transformation $t \rightarrow -i\tau$. 
Let a photon travel across the black hole horizon, then the radial geodesic is given by,
\begin{equation}
 \dot r=\frac{dr}{d\tau}=\pm i f(r)~,
\end{equation}
where the $+$ and $-$ sign correspond to the outgoing and ingoing paths, respectively.
Now the action given in equation (\ref{adiabatic covariant action}) can be written as,
\begin{equation}
\oint p_i d q_i=-4i\int_{r ^{out}} ^{r ^{in}} \int_{0} ^{H} \frac{dH^\prime}{f \left(r\right)} dr~.
\label{adiabatic invariant integral_stringy}
\end{equation}
Using the near horizon approximation, $f(r)$ can be Taylor expanded as,
\begin{equation} 
 f\left(r\right) =f\left(r\right)~\vline~_{r_h}+(r-r_{h})\frac{d f\left(r\right)}{dr}~\vline~_{r_h}+\cdots.
 \label{Taylor expansion}
\end{equation}
We also know that at the horizon $r=r_h$ there is a pole. To avoid this we consider a contour integral in such a way that 
the half loop is going above the pole from right to left. Evaluating the adiabatic invariant integral 
(\ref{adiabatic invariant integral_stringy}) using the Cauchy's integral theorem, we can arrive at,
\begin{equation}
 \oint p_i d q_i=4\pi \int_{0} ^{H} \frac{dH^\prime}{\kappa}=2 \hbar \int_{0} ^{H} \frac{dH^\prime}{T}~ ,
 \label{adiabatic invariant integral_temp}
\end{equation}
where $\kappa$ is the surface gravity of the black hole and it is related to the Hawking temperature by the 
relation $T=\frac{\hbar \kappa}{2\pi}$.

We can write the Smarr formula of the two dimensional dilatonic stringy black hole as,
\begin{equation}
 dM=dH=TdS- \phi dQ~.
\end{equation}
where $\phi$ is the electrostatic potential.
Then (\ref{adiabatic invariant integral_temp}) becomes,
\begin{equation}
 \oint p_i d q_i=2 \hbar \left( S- \frac{2 \pi m}{Q} (1-cos \alpha) \right)~.
 \label{adiabatic invariant integral_1}
\end{equation}
where $\alpha=sin^{-1}\left( q/m \right)$. Bohr-Sommerfeld quantization rule is given by, 
\begin{equation}
 \oint p_i d q_i=2\pi n \hbar, ~~~~~~~~ n=1,2,3,\cdots .
 \label{bohr sommerfeld quantization rule}
\end{equation}
Comparing the above two equations, (\ref{adiabatic invariant integral_1}) and (\ref{bohr sommerfeld quantization rule}), one can write the 
entropy spectrum as,
\begin{equation}
 S=n\pi+ \frac{2\pi m}{Q} \left( 1-cos \alpha \right)~.
 \label{entropy_spectra}
\end{equation}
Using equation (\ref{adm mass}), we can rewrite the entropy spectrum in terms of ADM mass of the black hole as,  
\begin{equation}
 S=n\pi+\frac{\pi}{Q^2} M e^{2\Phi_0} \left( 1-cos \alpha \right)~.
 \label{entropy_spectra_adm}
\end{equation}
It is interesting to note that the entropy of the two dimensional dilatonic stringy black holes is quantized.
From the above relation (\ref{entropy_spectra_adm}), it is evident that the entropy spectra depends on the value of the dilatonic field at the 
horizon, as $S\propto e^{2\Phi_0}$. Hence we may conclude that there is a background entropy due to this dilatonic field.

\section{Results and Discussion}
\label{conclusion}
In this work we have calculated the entropy of a $(1+1)$ dimensional dilatonic stringy black hole via the Euclidean treatment
of quantum gravity. We have studied the thermodynamics of the black hole. By calculating the temperature and heat capacity of
the black hole we conclude that the present black 
hole system does not undergo any kind of phase transition. This behaviour matches with the one dimensional Ising model system in statistical
studies \cite{stanley}.
We have investigated the quantization of entropy of $(1+1)$ dimensional dilatonic stringy black holes via adiabatic invariant integral 
method put forward by Majhi and Vagenas, as well as the Bohr-Sommerfeld quantization rule. We have found that the entropy spectrum  is quantized and 
also that the entropy spectrum depends on black hole parameters: electric charge, ADM mass and dilatonic field. This is supported by the area law in higher dimensional theory of gravity. By considering the two dimensional dilaton theory as the 
dimensional reduction from higher dimensional theories, one can conclude that the dilaton field is associated with the radius of the 
compactified coordinates. At this point, it is interesting enough to recall the ideas of inflationary cosmology \cite{veneziano,friedan,easther} which
imply the presence of a scalar field that drives the inflation. Here we can match this scalar field with the dilaton field. The dependence of entropy spectrum 
on dilatonic field points towards the microscopic origin of the black hole entropy and hence towards Quantum Gravity. This dilatonic field can also be 
considered as the source which drives the inflation in the context of inflationary cosmology. So further studies on these black hole solutions can unfold the mysteries regarding the
inflationary stage of the universe and also the microscopic origin of Bekenstein-Hawking entropy.

\section{Acknowledgments}
The authors wish to thank UGC, New Delhi for financial support through a major research project
sanctioned to VCK. VCK also wishes to acknowledge Associateship of IUCAA, Pune, India.

\end{document}